# Referenceless Proton Resonance Frequency Thermometry Using Deep Learning with Self-Attention

Authors Names and Degrees:

    Yueran Zhao, PhD Candidate[1],

    Chang-Sheng Mei, PhD[2,3],

    Nathan J. McDannold, PhD[4],

    Shenyan Zong, PhD[5],

    Guofeng Shen, PhD[1]

Author Affiliations:

    [1]Biomedical Instrument Institute, School of Biomedical Engineering, Shanghai Jiao Tong University, Shanghai, China

    [2]Department of Physics, Soochow University, Taipei, Taiwan

    [3]Department of Radiology, Harvard Medical School, Brigham and Women's Hospital, Boston, MA, USA

    [4]Department of Neurosurgery, Brigham and Women's Hospital, Boston, MA, USA

    [5]Department of Radiology, Shanghai Sixth People's Hospital Affiliated to Shanghai Jiao Tong University School of Medicine, Shanghai, China

Correspondence Author Info: Prof. Guofeng Shen, shenguofeng@sjtu.edu.cn and Dr. Shenyan Zong, shenyanzong@sjtu.edu.cn

Acknowledgements:

Grand Support:

Running Title: DL-Based Referenceless PRF Thermometry

# Referenceless Proton Resonance Frequency Thermometry Using Deep Learning with Self-Attention

**ABSTRACT**

**Background:** Accurate proton resonance frequency (PRF) MR thermometry is essential for monitoring temperature rise during thermal ablation with high intensity focused ultrasound (FUS). Conventional referenceless methods such as complex field estimation (CFE) and phase finite difference (PFD) tend to exhibit errors when susceptibility-induced phase discontinuities occur at tissue interfaces.

**Purpose:** To develop and evaluate a deep learning–based approach for referenceless PRF thermometry by reconstructing background complex MR images, and to compare its performance with established referenceless techniques.

**Study Type:** Retrospective single-center cohort study.

**Population:** 32 essential tremor patients (87 sonications, 1416 images). Data from 28 patients were used for model training and validation, and 4 independent patients (13 sonications, 146 images) formed the test set.

**Field Strength/Sequence:** 3T MRI system (Architect, GE Healthcare, Milwaukee, WI) using a spoiled gradient-echo (SPGR) sequence (TR/TE = 25.7/12.8 ms; matrix = 128×256 interpolated to 256×256; flip angle = 30°; FOV = 220×220 mm$^2$; bandwidth = ±5.81 kHz).

**Assessment:** Temperature estimation performance was evaluated by the investigators using voxel-wise error metrics in the heated region (MAE, Std, RMSE), 43°C isotherm delineation (Dice coefficient), and background stability in non-heated tissue. Agreement with reference PRF thermometry was assessed using Bland–Altman analysis and linear regression ($R^2$ and slope).

**Results:** The deep learning method achieved MAE = 0.64°C, Std = 0.80°C, RMSE = 0.82°C in heated regions, outperforming CFE (MAE = 1.74°C) and PFD (1.09°C). The 43°C isotherm Dice coefficient was 0.73 versus 0.49 for PFD. Background temperatures remained near zero (MAE = 0.20°C). Bland–Altman limits were −1.77°C to +1.37°C with bias −0.20°C; regression yielded $R^2$ = 0.99.

**DATA CONCLUSION:** The proposed deep learning approach improves accuracy and stability of referenceless PRF thermometry during transcranial FUS, with potential to enhance thermal monitoring safety and workflow efficiency.

**Evidence Level:** 3

**Technical Efficacy Stage:** 2

## 1. INTRODUCTION

Magnetic resonance thermometry (MRT) provides a non-invasive, non-ionizing method to monitor temperature changes during thermal ablation with high intensity focused ultrasound (FUS) [1, 2], radiofrequency [3, 4], microwave [5-7], and related modalities [8, 9]. By enabling real-time temperature mapping, MRT allows clinicians to visualize the course of thermal therapies and conduct procedures safely and effectively [10, 11]. In combination with FUS, the MR thermometry-guided techniques have been increasingly adopted for the treatment of essential tremors and tremor-dominated Parkinson's disease [12, 13].

Among MRT techniques, the proton resonance frequency (PRF) shift method is widely used for mapping temperature changes in aqueous tissues [14, 15]. At 3.0T, each 1°C increase produces an ~1.28 Hz shift in resonance frequency. Because this temperature dependence is encoded in the MR phase, conventional PRF thermometry derives temperature maps by subtracting a preheating (baseline) phase image from the phase acquired during heating. However, this baseline subtraction is highly vulnerable to inter-scan motion, making conventional PRF thermometry unreliable in many image-guided thermal therapy scenarios. In liver and breast FUS ablations, for example, respiration-related motion can introduce substantial phase mismatches between the baseline and heating scans, which in turn propagate into temperature error [16, 17].

To overcome the motion limitations of referenced PRF thermometry, Rieke et al. introduced referenceless PRF thermometry [18], which estimates the background phase in each frame by fitting a low-order polynomial to the unheated perimeter and extrapolating it into the heated region of interest (ROI), thereby avoiding a separate baseline and reducing sensitivity to inter-scan motion and frequency drift. Building on this framework, complex field estimation (CFE) [19] fits polynomials to the complex MR data (real and imaginary parts), implicitly enforcing phase smoothness and avoiding explicit unwrapping. Practical studies also clarified the geometry of the ROI and the surrounding region of reference (i.e., the unheated fitting band), as well as polynomial-order selection, to achieve stable reconstructions. Beyond CFE, variants differ mainly in how they model the background field and whether unwrapping is required. These include the near-harmonic (NH) [20] method, which models the background field based on (near) harmonic functions with appropriate boundary conditions and requires phase unwrapping. The phase-gradient (PG) [21] approach integrates smoothed local gradients, whereas the phase-finite-difference (PFD) [22] method fits discrete phase derivatives in the unheated region and then integrates them under boundary-continuity constraints. Comparative evaluations report that PFD and NH methods are the most robust among tested approaches [23], showing the lowest errors at low SNR and in rapidly varying backgrounds, while PG performs worst. However, error still increases for every method as SNR drops or susceptibility-induced phase discontinuities arise at tissue interfaces. In other words, PFD and NH are relatively more resilient rather than immune.

Deep learning has demonstrated strong performance across MRI, particularly in under-sampled reconstruction, noise suppression, and artifact reduction, where it frequently exceeds the capabilities of conventional compressed-sensing and parallel-imaging techniques [24, 25]. These advantages have motivated its use in PRF-based MR thermometry, mainly in the referenced setting, to accelerate temperature-map reconstruction, improve robustness at low SNR, and mitigate motion-related artifacts [26, 27]. In contrast, the application of deep learning to referenceless PRF thermometry remains far less developed. Recent reviews of motion-robust MR thermometry and deep-learning–based MRI reconstruction discuss deep learning primarily in the contexts of magnitude-image reconstruction, k-space acceleration, and motion correction, with limited attention to background-phase estimation for referenceless PRF thermometry [25, 26, 28-30]. Preliminary attempts using complex-valued networks with attention mechanisms [31] demonstrate the feasibility of applying deep learning to referenceless phase estimation, yet comprehensive frameworks are still lacking.

In this study, we propose a deep learning framework whose core is a complex-valued residual U-Net with self-attention termed C-SANet, to reconstruct background phase for referenceless PRF thermometry. The architecture learns spatial contextual relationships directly from complex MR images, mitigating the limitations of low-order polynomial models in the presence of anatomical complexity. A mixed training strategy that combines center-heated and background patches is used to enhance generalization. Network optimization employs a magnitude-phase decoupled loss to prioritize phase accuracy. C-SANet is evaluated on transcranial MR-guided focused ultrasound data and compared with representative referenced techniques. Results demonstrate improved accuracy, robustness, and spatial fidelity, supporting the clinical potential of deep learning-based referenceless MR thermometry.

## 2. MATERIALS AND METHODS

The study protocol was approved by the local ethics committee, and written informed consent was obtained from all patients prior to the procedure.

### 2.1 Referenced and Referenceless PRF thermometry

The PRF shift underpins MR thermometry [14]. PRF-based thermometry utilizes the linear dependence of the water proton resonance frequency on temperature, whereby local heating

produces a proportional phase shift in the MR signal. The resulting temperature change can be expressed as:

$$\Delta T = \frac{\phi - \phi_{ref}}{\alpha \cdot \gamma \cdot t_{TE} \cdot B_0} \quad [1]$$

where $\phi$ is the phase of the current image, $\phi_{ref}$ is the phase of the pre-heating reference image, $\alpha$ is the PRF temperature coefficient of water ($\approx -0.01\ ppm/°C$), is the gyromagnetic ratio of hydrogen, $t_{TE}$ is the echo time for a regular gradient echo sequence, and $B_0$ is the main field strength. While this reference-based formulation is accurate under controlled conditions, it is highly sensitive to inter-scan motion, which can introduce substantial temperature errors.

To address these limitations, referenceless PRF thermometry extends Equation 1 by dispensing with an explicit pre-heating reference [18]. It assumes heating is localized, while the surrounding, unheated tissue possesses a smoothly varying background phase that can act as implicit reference. Mathematically, this smooth-phase assumption is implemented by modeling the background phase within the unheated region of reference (ROR) as a low-order polynomial surface:

$$\phi_{bg}(x,y) = \sum_{i=0}^{m} \sum_{j=0}^{n} a_{ij} x^i y^j \quad [2]$$

where the coefficients $a_{ij}$ are estimated by fitting the measured phase within the ROR. The resulting background phase is extrapolated into the heated ROI and subtracted from the measured phase to estimate the temperature rise:

$$\Delta T(x,y) = \frac{\phi(x,y) - \phi_{bg}(x,y)}{\alpha \cdot \gamma \cdot t_{TE} \cdot B_0} \quad [3]$$

This referenceless formulation removes the need for pre-heating acquisitions and substantially reduces motion sensitivity. Its accuracy, however, hinges on the smooth-phase assumption and can deteriorate in regions of low SNR, pronounced susceptibility variations, or near complex tissue interfaces [23] – motivating data-driven approaches for more reliable background phase estimation, such as the deep learning method introduced here.

## 2.2 C-SANet for Background Complex Image Reconstruction

Building on the referenceless PRF thermometry framework, C-SANet is introduced to reconstruct the background complex image within heated regions. Unlike the conventional reference-based method depicted in Figure 1a, which requires baseline acquisition, C-SANet shown in Figure 1b operates in a completely referenceless manner. As outlined in Figure 1b, complex-valued image patches are cropped around ROIs, shown as red rectangles, and the heated region, indicated by yellow circles within each patch is masked. C-SANet internally processes these patches by decomposing the complex values into real and imaginary components to reconstruct the missing background image. The surrounding unheated tissue in each patch provides contextual information from which C-SANet, with architecture detailed in Figure 1c, restores the masked complex image. The reconstructed image is then used to compute the temperature map via the PRF shift model, preserving consistency with the theoretical formulation while eliminating the need for baseline images.

Conventional estimators such as low-order polynomials and finite-difference integration relay on spatial smoothness, often degrading under anatomical complex conditions. In contrast, C-SANet learns spatial regularities directly in the complex domain, enabling inference of the background image inside mask regions despite susceptibility variations. Operating on real-imaginary components also avoids explicit phase unwrapping, simplifying the pipeline and limiting error propagation.

As illustrated in Figure 1c, C-SANet adopts a complex-valued residual U-Net with integrated self-attention, comprising four down- and up-sampling stages connected by skips links to preserve multi-scale context. The self-attention module enables the network to adaptively weigh the importance of different spatial locations within the input feature maps. Specifically, from a given feature map, three distinct linear projections are derived: a query ($q$), a key ($k$), and a value ($v$). The attention mechanism is then formulated as:

$$Attn(q,k,v) = Softmax(\frac{q \otimes k}{\sqrt{c}}) \otimes v \quad [4]$$

where $c$ is the channel dimension of the key vectors. In our context, the query ($q$) corresponds to features at the location to be reconstructed within the masked ROI, while keys ($k$) and values ($v$) represent features from all surrounding context pixels. The scaled dot-product computes compatibility scores between $q$ and each $k$, generating a spatial attention map through softmax normalization. This map weights the $v$ vectors to aggregate the most relevant contextual information from the unheated region. This mechanism enables the network to dynamically focus on anatomically relevant surroundings while attenuating distant regions, aligning with the physical prior that background phase is locally correlated. Such context-aware attention enhances both reconstruction accuracy and model interpretability.

To emphasize phase precision, a magnitude-phase decoupled objective is used [32]:

$$loss_{dc} = d + \alpha \cdot l = ||\hat{y}| - |y|| + \alpha \cdot |\hat{y}| \cdot \mathbf{A}(\hat{y} \times \bar{y}) \quad [5]$$

where $y$ and $\hat{y}$ are the ground-truth and reconstructed complex images, $\bar{y}$ denotes complex conjugation, and $angle(\cdot)$ measures the angle discrepancy on the complex unit circle. The loss emphasizes accurate phase reconstruction for reliable PRF thermometry.

## 2.3 Training Data Preparation

This study analyzed clinical data from 32 essential tremor patients who underwent transcranial FUS ablation using the InSightec ExAblate 4000 system (Haifa, Israel), with temperature changes monitored using a 3T MRI system (Architect, GE Healthcare, Milwaukee, WI). The MRI acquisition was performed using a spoiled gradient echo (SPGR) sequence with TR/TE = 25.7/12.8 ms, matrix size = 128×256 interpolated to 256×256, flip angle = 30°, field of view (FOV) = 220×220 mm², and bandwidth = ±5.81 kHz. Axial temperature maps from 87 sonications were included, yielding 1416 images for the dataset.

As illustrated in Figure 2a, center-heated samples were formed by cropping square patches of size $N \times N$ around the heating center. Within each patch, a circular ROI of radius $r$ was masked to zero in the input, and corresponding complex image from the baseline frame served as ground truth. This setup trains C-SANet to reconstruct background complex image within heated regions using surrounding context. In total, 1270 center-heated patches were pre-extracted offline from 114 heating experiments across the 32 patients.

To improve generalization, the dataset was expanded online to include background-only samples so the model encounters both heating-induced patterns and broad anatomical variability. As shown in Figure 2b, background only samples were extracted from non-heated tissue: $N \times N$ patches were cropped, a central circular region was masked in the input, and the original complex image from the same frame served as the target. Each training batch mixed a fixed proportion of both patch types to balance reconstruction within heated ROIs with variability in surrounding anatomy.

Training solely on center-heated patches would bias the model toward central ROIs, whereas using only background-only patches would underrepresent thermal patterns. Accordingly, during training, a fixed quota of pre-extracted center-heated patches and dynamically resampled background-only patches, shown in Figure 2b, were assembled for each iteration, providing effectively unbounded background diversity. This hybrid schedule allows C-SANet to learn heating-induced image alterations while capturing whole-brain anatomical variability, improving accuracy and robustness to background field variability. By operating directly on complex-valued data, C-SANet leverages spatial continuity and avoids explicit phase unwrapping.

### 2.4 Training Configuration and Dataset Composition

C-SANet was trained with a mixture of center-heated and non-heated background patches , as shown in Figure 2c. Each iteration used a 70/30 split-70% center-heated to emphasize thermal-field patterns and 30% background to promote generalization to normal phase variations. This ratio was empirically determined to provide the optimal balance between reconstruction accuracy and generalization, as will be presented in the Results section. This configuration yielded stable convergence and mitigated bias toward heated regions.

Input were two-channel patches encoding the real and imaginary components of the complex MR image. The encoder-decoder employed feature widths of 32, 64, 128, and 256 with 3×3 convolutions throughout. Downsampling used 2D max pooling; unsampling used transposed convolutions. Each stage contained two residual convolutional blocks with skip connections to facilitate gradient propagation and stabilize training.

All models were implemented in PyTorch (version 2.4.1) and trained on NVIDIA A6000 GPUs using CUDA 12.4. We employed the AdamW optimizer [33] with an initial learning rate of $5×10^{-4}$, weight decay of $1×10^{-4}$, and cosine annealing scheduler. Models were trained for 700 epochs with a batch size of 10, following established hyperparameter guidelines from original publications with adjustments for our specific architecture.

### 2.5 Evaluation Protocol

Performance, robustness, and generalization were assessed on 13 transcranial heating experiments (4 subjects, 146 frames), spanning center-heated and peripheral non-heated regions. Each sonication included an unheated baseline complex image plus a time series of 10~16 heating frames. Ground-truth temperature per frame was computed via PRF shift from the phase difference relative to baseline, and accuracy was evaluated at a fixed heating center corresponding to the maximal thermal rise.

To investigate the influence of spatial context on reconstruction, we first examined the effects of crop size and ROI radius. We tested crop sizes of 27.5×27.5 mm² (1/64 FOV), 55×55 mm² (1/16 FOV), 110×110 mm² (1/4 FOV), and 220×220 mm² (full FOV), along with ROI radii of 6.0, 6.9, 7.7, and 8.6 mm, to characterize the trade-off between local detail and global context and to identify optimal input settings for temperature estimation in heated regions.

We further evaluated the sensitivity to training data balance by varying the sampling ratio between center-heated and background patches across five configurations (1:9, 3:7, 5:5, 7:3, 9:1), thereby probing the impact of thermal pattern diversity on reconstruction quality.

Finally, C-SANet was compared against conventional referenceless techniques: CFE and PFD. For CFE, a sixth-order 2D complex polynomial was fitted within the ROR, defined as a circular region surrounding but excluding the ROI. The configuration used an ROI radius of 1.7 mm and an ROR radius of 3.4 mm, with the fitted model extrapolated into the ROI to estimate the background complex signal and compute the phase difference. For PFD, phase gradients in the ROR in both x and y directions were fitted to a third-order 2D polynomial. The respective ROI and ROR radii were set to 7.7 mm and 10.3 mm, and the background phase inside the ROI was reconstructed by numerically integrating the fitted gradient field using a symmetric dual-boundary integration scheme.

Specificity in non-heated tissue was quantified through a background consistency analysis across all 146 frames. For each frame, 200 circular ROIs were randomly sampled from valid non-heated tissue located at least 17.2 mm from the heating center. The mean predicted temperature across these ROIs was calculated per frame, and global summary statistics, including mean and standard deviation, were aggregated over the entire dataset.

**2.6 Evaluation Metrics**

To comprehensively assess performance, multiple complementary metrics were used.

Voxel-wise accuracy and precision within the heated region were quantified by computing absolute temperature errors for each voxel and reporting the mean absolute error (MAE), standard deviation (Std), and root mean square error (RMSE).

Spatial consistency of the thermal distribution was evaluated by adapting the Dice coefficient to measure the overlap between predicted and reference regions exceeding the 43°C threshold. Although originally devised for segmentation, this usage assesses the correct delineation of clinically relevant thermal zones, as 43°C is a commonly cited boundary in mild hyperthermia.

Agreement between predicted and reference temperatures, along with potential systematic bias, was analyzed using Bland-Altman plots and linear regression.

Together, these metrics provide a rigorous view of both local estimation accuracy and global predictive reliability.

## 3. RESULTS

### 3.1 Optimizing Hyperparameters for Temperature Prediction

As summarized in Figure 3, a systematic analysis of input parameters was conducted to identify the optimal trade-off between contextual information and local precision. The corresponding heatmap visualizes model performance across a matrix of crop sizes, ranging from 1/16 of the FOV to the full 220×220 mm², and ROI radii, ranging from 6.0, 6.9, 7.7 and 8.6 mm, evaluated using a composite of metrics including MAE, Dice, Std and RMSE, where warmer colors indicate superior performance. The data reveal that very small crops, such as 27.5×27.5 mm² limited the use of broader anatomical context, whereas very large crops like the full FOV introduced redundancy that diluted feature relevance. Among all combinations, a crop size of 55×55 mm², accounting for 1/4 of the FOV, paired with an ROI radius of 7.7 mm provided the optimal balance, yielding the lowest Std and RMSE of 0.82°C each; this configuration was consequently adopted for all subsequent experiments.

### 3.2 Performance Variation with Training Sample Ratio

Training-set composition strongly affected generalization, as summarized in Table 1. A high proportion of center-heated patches at 0.9 delivered competitive accuracy within heated regions with an MAE of 0.62°C, but degraded background behavior with an MAE of 0.55°C, consistent with overfitting to thermal patterns. At the other extreme, a low ratio of 0.1 improved background stability yet reduced accuracy in heated regions, yielding an MAE of 0.8°C. A 0.7 center-heated ratio provided the best compromise for C-SANet, sustaining strong performance in heated regions with an MAE of 0.64°C and a Dice of 0.76, while achieving superior background specificity with an MAE of 0.2°C, an Std of 0.2°C and an RMSE of 0.24°C, as summarized in Table 2. This setting effectively balances learning of thermal-induced signal changes with exposure to diverse anatomical backgrounds.

### 3.3 Comparative analysis with CFE and PDF Methods

Representative results from one test case as shown in Figure 4. Figure 4a displays the predicted temperature map at the frame of peak heating. Figure 4b depicts the temporal temperature profile at the central heated voxel, demonstrating accurate tracking of heating dynamics. Figure 4c shows 43°C isotherms derived from predicted maps: the PFD contours exhibited noticeable misalignment and underestimation of the ablation zone, whereas C-SANet produces contours closely matching the true heated region. Because CFE relies on extrapolation within a small ROI, it failed to generate a complete 43°C isotherm and was therefore excluded from the Dice analysis.

Quantitative results across all test subjects shown in Table 3 indicate clear superiority of C-SANet. The mean absolute error was 0.64°C - an improvement of approximately 41% over PFD (1.09°C) and 63% over CFE (1.74°C). This accuracy gain was accompanied by higher precision, reflected in the lowest standard deviation (0.8°C) and RMSE (0.82°C). The ability to delineate the therapeutic ablation zone was further supported by an average Dice coefficient of 0.73 for the 43°C isotherm, nearly 50% higher than that of PFD (0.49).

Statistical agreement analyses corroborate these findings. Bland-Altman plots (Figure 5a) show the narrowest limits of agreement (-1.77°C to +1.37°C) and the smallest mean bias (-0.20°C) for C-SANet, indicating strong concordance with the reference without systematic error. Linear regression shown in Figure 5b, likewise yields a coefficient of determination ($R^2$) closest to 1 ($R^2$) and a slope nearest unity, underscoring high precision and reliability.

### 3.4 Background Region Consistency and Specificity

Evaluating non-heated background regions is essential for assessing robustness to false positives. C-SANet showed strong stability, with temperature estimates tightly clustered around 0°C, exhibiting an MAE of 0.2°C and an Std of 0.2°C, as detailed in Table 4. As illustrated in Figure 6a~b, its background estimates were consistently centered at zero with minimal dispersion, in clear contrast to the pronounced bias and variability observed for both CFE and PFD. These results indicate effective suppression of spurious temperature elevations in unheated tissue – an attribute critical for clinical safety.

## 4. DISCUSSION

This study introduced C-SANet , a deep learning-based framework for referenceless PRF MR thermometry that reconstructs background complex images within heated regions, yielding accurate temperature estimates and precise localization of thermal elevation. Mean error and standard deviation in heated regions were low, and the 43°C isotherm achieved a high Dice coefficient. Relative to conventional CFE and PFD, performance remained superior under background field, underscoring the value of data-driven background modeling.

The advantages of C-SANet relative to existing referenceless techniques stem from differences in modeling capacity. CFE and PFD rely on polynomial or derivative-based approximations and are therefore constrained by assumptions of local smoothness. These assumptions can be violated near susceptibility interfaces. In contrast, C-SANet learns spatial dependencies directly in the complex domain, allowing it to capture patterns that analytic models cannot represent and to restore background phase without explicit unwrapping. The mixed training strategy—combining center-heated and background-only patches—also contributed to the robustness of the framework. This design prevents overfitting to specific heating geometries while exposing the model to a wider range of anatomical variability. As a result, C-SANet is better equipped to handle both thermally perturbed and unperturbed regions, supporting stable reconstruction under diverse background conditions.

Several limitations should be acknowledged. Although referenceless PRF thermometry is well suited for anatomies with substantial physiological motion because it does not require a baseline, the present study did not include experiments in highly dynamic organs such as the liver or heart. Future work will extend the evaluation to these motion-prone regions to more comprehensively characterize performance under different anatomical and motion conditions. In addition, the current network estimates temperature on a frame-by-frame basis without explicitly incorporating temporal information. Incorporating temporal consistency constraints—such as recurrent, temporal-convolutional, or transformer-based modules [33-35]—represents a natural direction for further strengthening stability, particularly in low-SNR or rapidly varying background-field settings. Broader validation across additional orientations, organ systems, and heating strategies will be pursued to support translation into diverse clinical scenarios.

## 5. CONCLUSION

This study presents C-SANet, a deep learning framework for referenceless PRF thermometry based on a complex-valued residual U-Net with integrated self-attention. By modeling multi-scale spatial structure and contextual dependencies directly in the complex domain, C-SANet accurately reconstructs background phase and yields robust, precise temperature estimates. Trained with a mix scheme that couples local heating and background context, C-SANet moves toward real-time, reference-free monitoring and has the potential to streamline MR-guided focused ultrasound workflows and enhance safety. These results establish a strong baseline and paves the way for future clinical applications that require robust, referenceless thermometry.

## DATA AVAILABILITY STATEMENT

The source code for the proposed referenceless MR Thermometry network C-SANet is publicly available at: https://github.com/hiiamrr/Referenceless-Proton-Resonance-Frequency-Thermometry-Using-Deep-Learning-with-Self-Attention.git

**TABLE 1.** Quantitative performance in heated regions across training data compositions.

| Center Ratio | MAE(°C) | Std(°C) | RMSE(°C) | Prct98(°C) | Prct2(°C) | Dice (≥43°C) |
|---|---|---|---|---|---|---|
| 0.9 | **0.62** | **0.79** | 0.84 | 1.07 | -0.51 | 0.75 |
| 0.7 | 0.64 | 0.80 | **0.82** | **1.00** | -0.60 | 0.76 |
| 0.5 | 0.66 | 0.83 | 0.85 | 1.01 | -0.64 | 0.75 |
| 0.3 | 0.68 | 0.84 | 0.87 | 1.06 | -0.62 | 0.74 |
| 0.1 | 0.80 | 0.89 | 0.99 | 1.32 | **-0.46** | **0.77** |

Abbreviations: MAE, mean absolute error; Std, standard deviation; RMSE, root mean square error; Prct98, 98th percentile; Prct2, 2nd percentile; Dice, dice similarity coefficient.

**TABLE 2.** Quantitative background region performance across training data compositions.

| Center Ratio | MAE(°C) | Std(°C) | RMSE(°C) | Prct98(°C) | Prct2(°C) |
|---|---|---|---|---|---|
| 0.9 | 0.55 | 0.52 | 0.74 | 0.49 | -1.54 |
| 0.7 | **0.20** | 0.20 | **0.24** | 0.24 | **-0.53** |
| 0.5 | 0.29 | 0.23 | 0.35 | 0.18 | -0.72 |
| 0.3 | 0.22 | 0.24 | 0.30 | 0.31 | -0.65 |
| 0.1 | 0.23 | **0.18** | 0.27 | **0.14** | -0.56 |

Abbreviations: MAE, mean absolute error; Std, standard deviation; RMSE, root mean square error; Prct98, 98th percentile; Prct2, 2nd percentile.

**TABLE 3.** Quantitative comparison of temperature prediction methods in heated regions.

| Methods | MAE(°C) | Std(°C) | RMSE(°C) | Prct98(°C) | Prct2(°C) | Dice (≥43°C) |
|---|---|---|---|---|---|---|
| C-SANet | **0.64** | **0.80** | **0.82** | 1.77 | **-1.36** | **0.73** |
| CFE | 1.74 | 2.20 | 2.62 | 2.87 | -5.74 | - |
| PFD | 1.09 | 1.16 | 1.37 | **1.56** | -3.00 | 0.49 |

Abbreviations: MAE, mean absolute error; Std, standard deviation; RMSE, root mean square error; Prct98, 98th percentile; Prct2, 2nd percentile; Dice, dice similarity coefficient; C-SANet, complex-valued self-attention network; CFE, complex field estimation; PFD, phase finite difference.

**TABLE 4.** Quantitative background estimation performance under different models.

| Methods | MAE(°C) | Std(°C) | RMSE(°C) | Prct98(°C) | Prct2(°C) |
|---|---|---|---|---|---|
| C-SANet | **0.20** | **0.20** | **0.24** | **0.24** | **-0.53** |
| CFE | 1.22 | 1.73 | 1.73 | 3.48 | -3.29 |
| PFD | 1.126 | 1.50 | 1.50 | 2.92 | -2.95 |

Abbreviations: MAE, mean absolute error; Std, standard deviation; RMSE, root mean square error; Prct98, 98th percentile; Prct2, 2nd percentile; C-SANet, complex-valued self-attention network; CFE, complex field estimation; PFD, phase finite difference.

**FIGURE LEGENDS**

**FIGURE 1.** Schematic diagrams of MR thermometry frameworks. (a) presents conventional baseline-referenced PRF thermometry workflow. (b) shows the processing pipeline, starting from complex MR input images, followed by ROI masking, patch extraction, network inference to predict the missing background phase, and temperature calculation. (c) presents the architecture of C-SANet with four-level encoder–decoder and skip connections, with the detailed designs of convolutional blocks, residual convolutional blocks, and attention blocks, illustrated at the bottom.

**FIGURE 2.** Construction of the training dataset. (a) illustrates the preparation of center-heated training samples, where a N×N patch is cropped around the heating center, the central ROI is masked as network input, and the target is constructed by filling the masked ROI with the corresponding region from the first unheated frame. (b) shows the generation of background non-heated training samples, where central circular regions (radius = 20) from each frame are masked from each frame to eliminate heated areas, followed by random sampling ROIs within the remaining tissue. The intact patch itself serves as the target. (c) depicts the final training dataset obtained by fusing the two sample types in each batch with a fixed ratio of 0.7 (center-heated) to 0.3 (background-nonheated). This strategy ensures the model can generalize to both heated ROIs near the center and background ROIs at the tissue periphery.

**FIGURE 3.** Heatmap of temperature prediction MAE, Dice, Std, RMSE, in heated regions under varying combinations of crop size and ROI radius. Red indicates lower MAE, Std, RMSE and higher Dice. The result shows optimal performance is achieved at crop size of 55×55 mm$^2$ and ROI radius of 7.7 mm.

**FIGURE 4.** Representative temperature estimation results from one test case. (a) Temperature maps at the peak heating frame reconstructed by ground truth (GT) and three comparative methods: DL, CFE, and PFD. (b) Temporal evolution of the central heated pixel. The DL method closely tracks the GT trajectory, while CFE underestimates the heating peak and PFD shows inferior performance. (c) Comparison of 43°C isothermal contours. The DL contours align well with GT, whereas PFD exhibits noticeable deviations. The CFE method failed to generate a complete 43°C isotherm due to its reliance on extrapolation within a small ROI and was therefore excluded from this analysis.

**FIGURE 5.** Bland–Altman and regression analyses of temperature estimation. (a) shows Bland–Altman plots comparing predicted and ground-truth temperatures across all test cases for C-SANet, CFE, and PFD methods. C-SANet demonstrates the narrowest limits of agreement and the smallest bias, while CFE and PFD exhibit wider variability. (b) presents linear regression analysis between predicted and ground-truth temperatures. C-SANet achieves the highest coefficient

of determination (R²) and a slope closest to unity, indicating superior accuracy and reliability compared with classical approaches.

**FIGURE 6.** Background temperature of predicted versus reference temperatures (per frame, averaged over 200 ROIs) for C-SANet, CFE, and PFD. (a) shows estimated temperature over time in randomly sampled background regions. C-SANet maintains predictions close to 0°C, indicating high robustness and specificity. (b) shows distribution of background temperature estimates across frames for different training strategies. C-SANet demonstrates lower variance and reduced outliers.

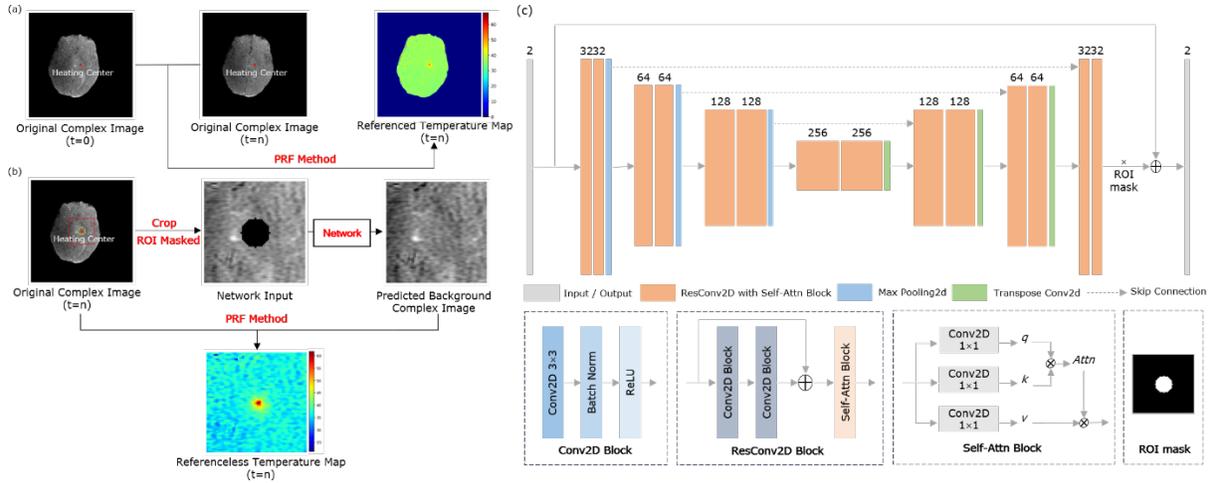

**FIGURE 1.** Schematic diagrams of MR thermometry frameworks. (a) presents conventional baseline-referenced PRF thermometry workflow. (b) shows the processing pipeline, starting from complex MR input images, followed by ROI masking, patch extraction, network inference to predict the missing background phase, and temperature calculation. (c) presents the architecture of C-SANet with four-level encoder–decoder and skip connections, with the detailed designs of convolutional blocks, residual convolutional blocks, and attention blocks, illustrated at the bottom.

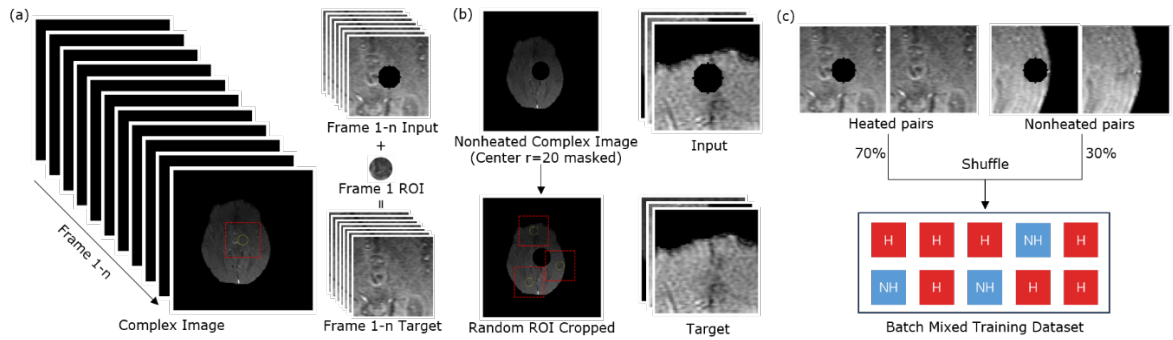

**FIGURE 2.** Construction of the training dataset. (a) illustrates the preparation of center-heated training samples, where a N×N patch is cropped around the heating center, the central ROI is masked as network input, and the target is constructed by filling the masked ROI with the corresponding region from the first unheated frame. (b) shows the generation of background non-heated training samples, where central circular regions (radius = 20) from each frame are masked from each frame to eliminate heated areas, followed by random sampling ROIs within the remaining tissue. The intact patch itself serves as the target. (c) depicts the final training dataset obtained by fusing the two sample types in each mini-batch with a fixed ratio of 0.7 (center-heated) to 0.3 (background-nonheated). This strategy ensures the model can generalize to both heated ROIs near the center and background ROIs at the tissue periphery.

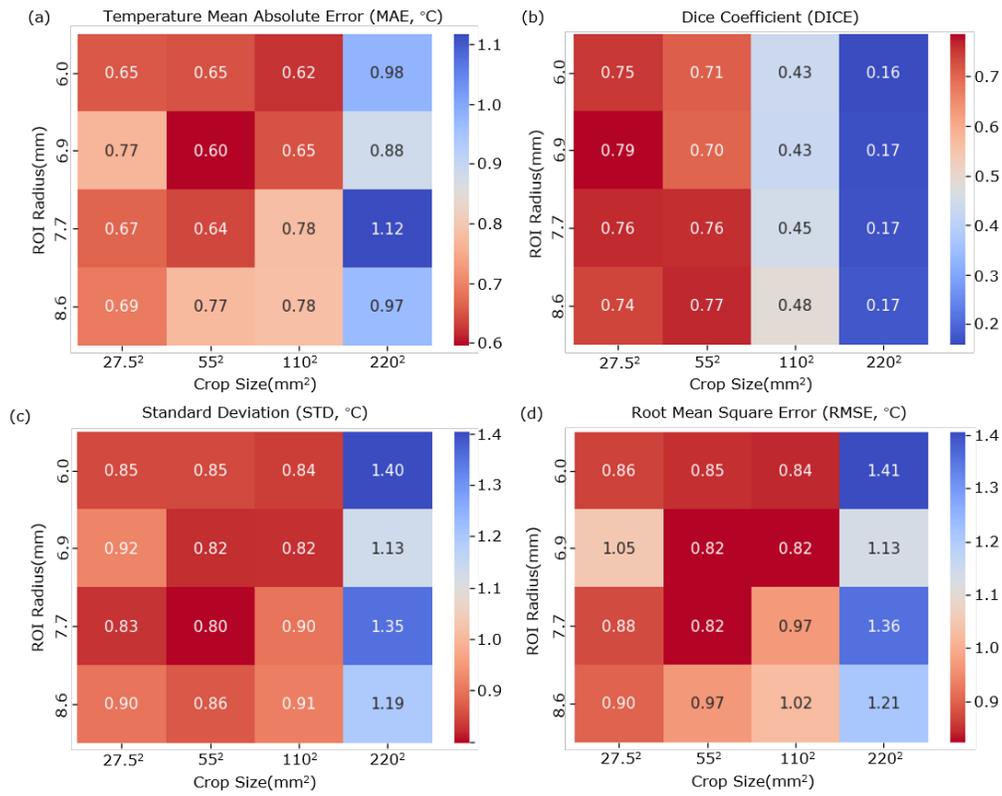

**FIGURE 3.** Heatmap of temperature prediction MAE, DICE, Std, RMSE, in heated regions under varying combinations of crop size and ROI radius. Red indicates lower MAE, Std, RMSE and higher DICE. The result shows optimal performance is achieved at crop size of 55×55 mm$^2$ and ROI radius of 7.7 mm.

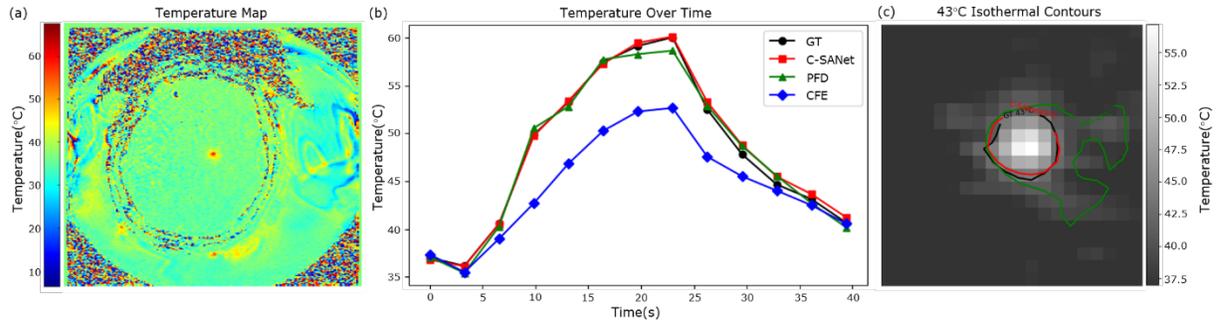

**FIGURE 4.** Representative temperature estimation results from one test case. (a) Temperature maps at the peak heating frame reconstructed by ground truth (GT) and three comparative methods: DL, CFE, and PFD. (b) Temporal evolution of the central heated pixel. The DL method closely tracks the GT trajectory, while CFE underestimates the heating peak and PFD shows inferior performance. (c) Comparison of 43°C isothermal contours. The DL contours align well with GT, whereas PFD exhibits noticeable deviations. The CFE method failed to generate a complete 43°C isotherm due to its reliance on extrapolation within a small ROI and was therefore excluded from this analysis.

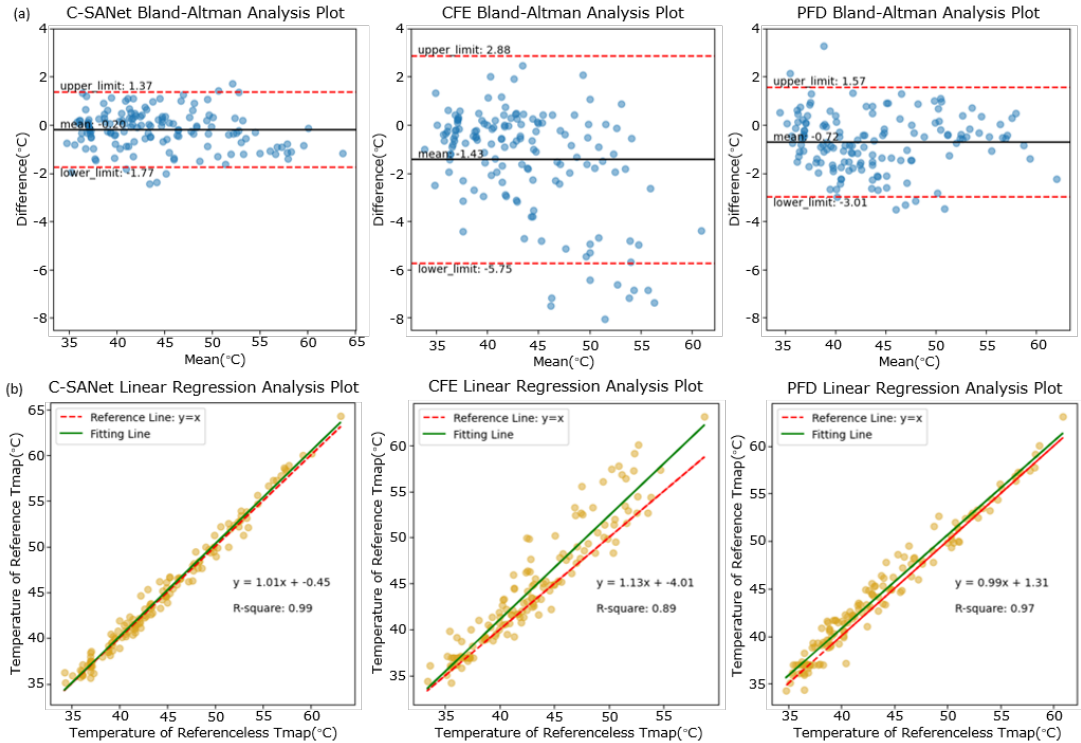

**FIGURE 5.** Bland–Altman and regression analyses of temperature estimation. (a) shows Bland–Altman plots comparing predicted and ground-truth temperatures across all test cases for C-SANet, CFE, and PFD methods. C-SANet demonstrates the narrowest limits of agreement and the smallest bias, while CFE and PFD exhibit wider variability. (b) presents linear regression analysis between predicted and ground-truth temperatures. C-SANet achieves the highest coefficient of determination ($R^2$) and a slope closest to unity, indicating superior accuracy and reliability compared with classical approaches.

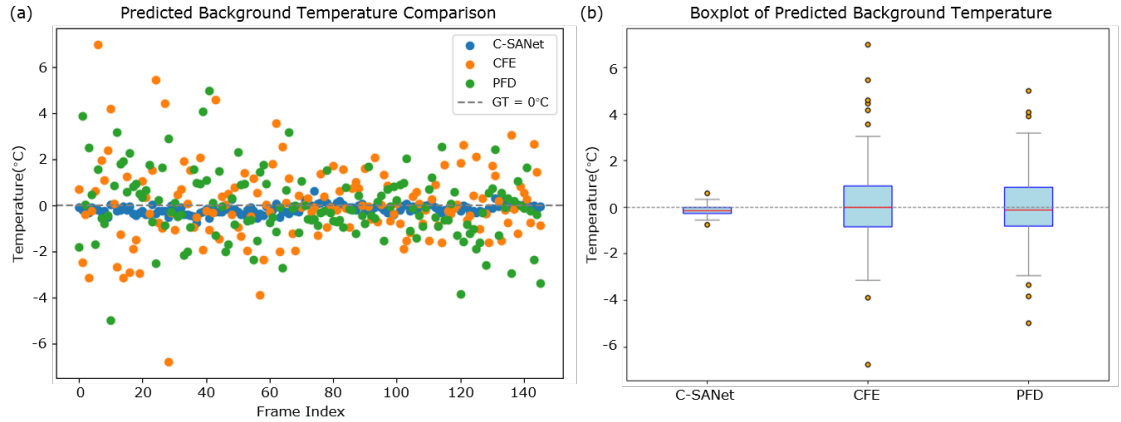

**FIGURE 6.** Background temperature of predicted versus reference temperatures (per frame, averaged over 200 ROIs) for C-SANet, CFE, and PFD. (a) shows estimated temperature over time in randomly sampled background regions. C-SANet maintains predictions close to 0°C, indicating high robustness and specificity. (b) shows distribution of background temperature estimates across frames for different training strategies. C-SANet demonstrates lower variance and reduced outliers.